\documentclass{ws-p8-50x6-00}

\newcommand{\Tr}{{\rm Tr}}

\begin{document}

\title{The $U(1)$ axial symmetry and the chiral transition in QCD}

\author{Enrico Meggiolaro}

\address{Dipartimento di Fisica ``E. Fermi'', Universit\`a di Pisa,\\
Via Buonarroti 2, I--56127 Pisa, Italy\\
E-mail: enrico.meggiolaro@df.unipi.it}


\maketitle

\abstracts{
We discuss the role of the $U(1)$ axial symmetry for the phase structure of
QCD at finite temperature. We expect that, above a certain critical
temperature, also the $U(1)$ axial symmetry will be restored.
We will try to see if this transition has (or has not) anything to do
with the usual chiral transition: various possible scenarios are discussed.
In particular, we analyse a scenario in which the $U(1)$ axial symmetry is
still broken above the chiral transitions. We will show that this scenario
can be consistently reproduced using an effective Lagrangian model.
A new order parameter is introduced for the $U(1)$ axial symmetry.}

\section{Introduction}

It is generally believed that a phase transition which occurs in QCD at a
finite temperature is the restoration of the spontaneously broken 
$SU(L) \otimes SU(L)$ chiral symmetry in association with $L$ massless quarks.
At zero temperature the chiral symmetry is broken spontaneously by the
condensation of $q\bar{q}$ pairs and the $L^2-1$ $J^P=0^-$ mesons
are just the Nambu--Goldstone (NG) bosons associated with this breaking.
At high temperatures the thermal energy breaks up the $q\bar{q}$ condensate,
leading to the restoration of chiral symmetry. We expect that this property
not only holds for massless quarks but also continues for a small mass region.
The order parameter for the chiral symmetry breaking is apparently 
$\langle \bar{q}q \rangle \equiv \sum_{i=1}^L \langle \bar{q}_i q_i \rangle$: 
the chiral symmetry breaking corresponds to the non--vanishing of 
$\langle \bar{q}q \rangle$ in the chiral limit $sup(m_i) \to 0$.
From lattice determinations of the chiral order parameter
$\langle \bar{q}q \rangle$ one knows that the $SU(L) \otimes SU(L)$ chiral
phase transition temperature $T_{ch}$, defined as the temperature at which
the chiral condensate $\langle \bar{q}q \rangle$ goes to zero (in the chiral
limit $sup(m_i) \to 0$), is nearly equal to the deconfining temperature
$T_c$~(see, e.g., Ref.~\cite{Blum-et-al.95}).
But this is not the whole story: QCD possesses not only an 
approximate $SU(L) \otimes SU(L)$ chiral symmetry, for $L$ light quark 
flavours, but also a $U(1)$ axial symmetry (at least at the classical 
level). The role of the $U(1)$ symmetry for the finite temperature phase 
structure has been so far not well studied and it is still an open question
of hadronic physics whether the fate of the $U(1)$ chiral symmetry of QCD has
or has not something to do with the fate of the $SU(L) \otimes SU(L)$ chiral
symmetry. In the following sections we shall try to answer these questions:
\begin{itemize}
\item{} At which temperature is the $U(1)$ axial symmetry restored?
(if such a critical temperature does exist!)
\item{} Does this temperature coincide with the deconfinement temperature 
and with the temperature at which the $SU(L)$ chiral symmetry is restored?
\end{itemize}

\section{Topological and chiral susceptibilities}

In the ``Witten--Veneziano mechanism''~\cite{Witten79,Veneziano79}
for the resolution of the $U(1)$ problem, a fundamental role is played by
the so--called ``topological susceptibility'' in a QCD without
quarks, i.e., in a pure Yang--Mills (YM) theory, in the large--$N_c$ limit
($N_c$ being the number of colours):
\be
A = \displaystyle\lim_{k \to 0}
\displaystyle\lim_{N_c \to \infty}
\left\{ -i \displaystyle\int d^4 x e^{ikx} \langle T Q(x) Q(0) \rangle
\right\} ,
\label{eqn1}
\ee
where $Q(x) = {g^2 \over 64\pi^2}\varepsilon^{\mu\nu\rho\sigma} F^a_{\mu\nu}
F^a_{\rho\sigma}$ is the so--called ``topological charge density''.
This quantity enters into the expression for the squared mass of the $\eta'$:
$m^2_{\eta'} = {2L A \over F^2_\pi}$, where $L$ is the number of light quark
flavours taken into account in the chiral limit.
Therefore, in order to study the role of the $U(1)$ axial symmetry for the
full theory at non--zero temperatures, one should consider the YM topological
susceptibility $A(T)$ at a given temperature $T$, formally defined as in
Eq. (\ref{eqn1}), where now $\langle \ldots \rangle$ stands for the
expectation value in the full theory at a given temperature $T$.~\cite{EM1998}
The problem of studying the behaviour of $A(T)$ as a function of the
temperature $T$ was first addressed, in lattice QCD,
in Refs.~\cite{Teper86,EM1992}.
Recent lattice results~\cite{Alles-et-al.97} (obtained for the $SU(3)$
pure--gauge theory) show that the YM topological susceptibility $A(T)$
is approximately constant up to the critical temperature $T_c \simeq T_{ch}$,
it has a sharp decrease above the transition, but it remains different
from zero up to $\sim 1.2~T_c$.
In the Witten--Veneziano mechanism, a (no matter how small!)
value different from zero for $A$ is related to the breaking of 
the $U(1)$ axial symmetry, since it implies the existence of a 
pseudo--Goldstone particle with the same quantum numbers of the $\eta'$.
Therefore, the available lattice results show that the $U(1)$ chiral
symmetry is restored at a temperature $T_{U(1)}$ greater than $T_{ch}$.

Another way to address the same question is to look at the behaviour at
non--zero temperatures of the susceptibilities related to the
propagators for the following meson channels~\cite{Shuryak94}
(we consider for simplicity the case of $L=2$ light flavours):
the isoscalar $I=0$ scalar channel $\sigma$ (also known as $f_0$ in the
modern language of hadron spectroscopy), interpolated by the operator
$O_\sigma = \bar{q} q$;
the isovector $I=1$ scalar channel $\delta$
(also known as $a_0$), interpolated by the operator
$\vec{O}_\delta = \bar{q} {\vec{\tau} \over 2} q$;
the isovector $I=1$ pseudoscalar channel $\pi$, interpolated by the operator
$\vec{O}_\pi = i\bar{q} \gamma_5 {\vec{\tau} \over 2} q$;
the isoscalar $I=0$ pseudoscalar channel $\eta'$, interpolated by the operator
$O_{\eta'} = i\bar{q} \gamma_5 q$.
Under $SU(2)_A$ transformations, $\sigma$ is mixed with $\pi$: thus the
restoration of this symmetry at $T_{ch}$ requires identical correlators
for these two channels. Another $SU(2)$ chiral multiplet is $(\delta,\eta')$.
On the contrary, under the $U(1)_A$ transformations, $\pi$ is mixed
with $\delta$: so, a ``practical restoration'' of the $U(1)$ axial
symmetry should imply that these two channels become degenerate, with
identical correlators. Another $U(1)$ chiral multiplet is $(\sigma,\eta')$.
(Clearly, if both chiral symmetries are restored, then all $\pi$, $\eta'$,
$\sigma$ and $\delta$ correlators should become the same.)
In practice, one can construct, for each meson channel $f$, the
corresponding chiral susceptibility
\be
\chi_f = \displaystyle\int d^4x \langle O_f(x) O_f^\dagger(0) \rangle ,
\label{eqn2}
\ee
and then define two order parameters:
$\chi_{SU(2) \otimes SU(2)} \equiv \chi_\sigma - \chi_\pi$, and
$\chi_{U(1)} \equiv \chi_\delta - \chi_\pi$.
If an order parameter is non--zero in the chiral limit, then the
corresponding symmetry is broken.
Present lattice data for these quantities seem to indicate that the $U(1)$
axial symmetry is still broken above $T_{ch}$, up to $\sim 1.2~T_{ch}$,
where the $\delta$--$\pi$ splitting is small but still
different from zero.~\cite{Bernard-et-al.97,Karsch00,Vranas00}
In terms of the left--handed and right--handed quark fields
[$q_{L,R} \equiv {1 \over 2} (1 \pm \gamma_5) q$], one has the following
expression for the difference between the correlators for
the $\delta^+$ and $\pi^+$ channels:
\bea
\lefteqn{
{\cal D}_{U(1)}(x) \equiv \langle O_{\delta^+}(x) O_{\delta^+}^\dagger(0)
\rangle - \langle O_{\pi^+}(x) O_{\pi^+}^\dagger(0) \rangle } \nonumber \\
& & = 2 \left[ \langle \bar{u}_R d_L(x) \cdot \bar{d}_R u_L(0) \rangle
+ \langle \bar{u}_L d_R(x) \cdot \bar{d}_L u_R(0) \rangle \right] .
\label{eqn3}
\eea
(The integral of this quantity, $\int d^4x {\cal D}_{U(1)}(x)$, is just equal
to the $U(1)$ chiral parameter $\chi_{U(1)} = \chi_\delta - \chi_\pi$.)
What happens below and above $T_{ch}$?
Below $T_{ch}$, in the chiral limit $sup(m_i) \to 0$, the left--handed
and right--handed components of a given light quark flavour ({\it up} or
{\it down}, in our case with $L=2$) can be connected through the
quark condensate, giving rise to a non--zero contribution to the
quantity ${\cal D}_{U(1)}(x)$ in Eq. (\ref{eqn3}) (i.e., to the quantity
$\chi_{U(1)}$).
But above $T_{ch}$ the quark condensate is zero: so, how can the quantity
${\cal D}_{U(1)}(x)$ (i.e., the quantity $\chi_{U(1)}$) be different from zero
also above $T_{ch}$, as indicated by present lattice data?
The only possibility in order to solve this puzzle seems to be that of
requiring the existence of a genuine four--fermion local condensate,
which is an order parameter for the $U(1)$ axial symmetry and which
remains different from zero also above $T_{ch}$.
This will be discussed in section 4.

\section{Which chiral symmetry is restored in hot QCD?}

Let us define the following temperatures:
\begin{itemize}
\item{} $T_{ch}$: the temperature at which the chiral condensate
$\langle \bar{q} q \rangle$ goes to zero. The chiral symmetry
$SU(L) \otimes SU(L)$ is spontaneously broken below $T_{ch}$ and
it is restored above $T_{ch}$.
\item{} $T_{U(1)}$: the temperature at which the $U(1)$ axial symmetry
is (approximately) restored.
If $\langle \bar{q} q \rangle \ne 0$ also the $U(1)$ axial symmetry is
broken, i.e., the chiral condensate is an order parameter also for
the $U(1)$ axial symmetry. Therefore we must have: $T_{U(1)} \ge T_{ch}$.
\item{} $T_\chi$: the temperature at which the pure--gauge topological
susceptibility $A$ (approximately) drops to zero. Present lattice results
indicate that $T_\chi \ge T_{ch}$.~\cite{Alles-et-al.97}
Moreover, the Witten--Veneziano mechanism implies that $T_{U(1)} \ge T_\chi$.
\end{itemize}
The following scenario, that we shall call ``\underline{SCENARIO 1}'',
in which $T_\chi < T_{ch}$, is, therefore, immediately ruled out.
In this case, in the range of temperatures between $T_\chi$ and $T_{ch}$
the $U(1)$ axial symmetry is still broken by the chiral condensate, but the
anomaly effects are absent. In other words, in this range of temperatures
the $U(1)$ axial symmetry is spontaneously broken ({\it \`a la} Goldstone)
and the $\eta'$ is the corresponding NG boson, i.e., it is massless in the
chiral limit $sup(m_i) \to 0$, or, at least, as light as the pion $\pi$,
when including the quark masses.
This scenario was first discussed (and indeed really supported!) in
Ref.~\cite{Pisarski-Wilczek84}.
It is known that the $U(1)$ chiral anomaly effects are connected with
instantons.~\cite{tHooft76} It is also known that at high temperature $T$
the instanton--induced amplitudes are suppressed by the so--called
``Pisarski--Yaffe suppression factor'', due to the Debye--type screening:
\be
d{\cal N}_{inst}(T) \sim d{\cal N}_{inst}(T=0) \cdot
\exp \left[ -\pi^2 \rho^2 T^2 \left( {2N_c + L \over 3} \right) \right] ,
\label{eqn4}
\ee
$\rho$ being the instanton radius.
The argument of Pisarski and Wilczek in Ref.~\cite{Pisarski-Wilczek84} was
the following: {``If instantons themselves are the primary 
chiral--symmetry--breaking mechanism, then it is very difficult to
imagine the unsuppressed $U(1)_A$ amplitude at $T_{ch}$.''}
So, what was wrong in their argument?
The problem is that Eq. (\ref{eqn4}) can be applied only in the quark--gluon
plasma phase, since the Debye screening is absent below $T_{ch}$.
Indeed, Eq. (\ref{eqn4}) is applicable only for $T \ge 2T_{ch}$ and
one finds instanton suppression by at least two orders of magnitude
at $T \simeq 2T_{ch}$ (see Ref.~\cite{Shuryak94} and references therein).
Moreover, the qualitative picture of instanton--driven chiral symmetry
restoration which is nowadays accepted has significantly changed since
the days of Ref.~\cite{Pisarski-Wilczek84}.
It is now believed (see Ref.~\cite{Shuryak94} and references therein)
that the suppression of instantons is not the only way to ``kill'' the
quark condensate. Not only the number of instantons is important, but
also their relative positions and orientations. Above $T_{ch}$,
instantons and anti-instantons can be rearranged into some finite
clusters with zero topological charge, such as well--formed
``instanton--anti-instanton molecules''.

Therefore, we are left essentially with the two following scenarios.\\
\underline{SCENARIO 2}: $T_{ch} \le T_{U(1)}$, with
$T_{ch} \sim T_\chi \sim T_{U(1)}$.
If $T_{ch} = T_\chi = T_{U(1)}$, then, in the case of $L=2$ light flavours,
the restored symmetry across the transition is $U(1)_A \otimes SU(2)_L \otimes
SU(2)_R \sim O(2) \otimes O(4)$, which may yield a first--order phase
transition (see, for example, Ref.~\cite{Kharzeev-et-al.98}).\\
\underline{SCENARIO 3}: $T_{ch} \ll T_{U(1)}$, that is, the complete
$U(L)_L \otimes U(L)_R$ chiral symmetry is restored only well inside the
quark--gluon plasma domain.
In the case of $L=2$ light flavours, we then have at $T=T_{ch}$ the
restoration of $SU(2)_L \otimes SU(2)_R \sim O(4)$.
Therefore, we can have a second--order phase transition with the
$O(4)$ critical exponents. $L=2$ QCD at $T \simeq T_{ch}$ and the $O(4)$
spin system should belong to the same universality class.
An effective Lagrangian describing the softest modes is essentially
the Gell-Mann--Levy linear sigma model, the same as for the $O(4)$
spin systems (see Ref.~\cite{Pisarski-Wilczek84}).
If this scenario is true, one should find the $O(4)$ critical indices
for the quark condensate and the specific heat:
$\langle \bar{q} q \rangle \sim |(T - T_{ch})/T_{ch}|^{0.38 \pm 0.01}$,
and $C(T) \sim |(T - T_{ch})/T_{ch}|^{0.19 \pm 0.06}$.
Present lattice data partially support these results.

\section{The $U(1)$ chiral order parameter}

We make the assumption (discussed in the previous sections) that the $U(1)$ 
chiral symmetry is broken independently from the $SU(L) \otimes SU(L)$ 
symmetry. The usual chiral order parameter $\langle \bar{q} q \rangle$
is an order parameter both for $SU(L) \otimes SU(L)$ and for $U(1)_A$:
when it is different from zero, $SU(L) \otimes SU(L)$ is broken down to
$SU(L)_V$ and also $U(1)_A$ is broken.
Thus we need another quantity which could be an order parameter only for 
the $U(1)$ chiral symmetry.~\cite{EM1994,EM1995}
The most simple quantity of this kind was found by 'tHooft
in Ref.~\cite{tHooft76}.
For a theory with $L$ light quark flavours, it is a $2L$--fermion interaction
that has the chiral transformation properties of:
\be
{\cal L}_{eff} \sim \displaystyle{{\det_{st}}(\bar{q}_{sL}q_{tR})
+ {\det_{st}}(\bar{q}_{sR}q_{tL}) },
\label{eqn5}
\ee
where $s,t = 1, \ldots ,L$ are flavour indices, but the colour indices are 
arranged in a more general way  (see below).
It is easy to verify that ${\cal L}_{eff}$ is invariant under
$SU(L) \otimes SU(L) \otimes U(1)_V$, while it is not invariant under $U(1)_A$.
To obtain an order parameter for the $U(1)$ chiral symmetry, one can 
simply take the vacuum expectation value of ${\cal L}_{eff}$:
$C_{U(1)} = \langle {\cal L}_{eff} \rangle$.
The arbitrarity in the arrangement of the colour indices can be removed if we 
require that the new $U(1)$ chiral condensate is ``independent'' of the
usual chiral condensate $\langle \bar{q} q \rangle$, as explained in
Refs.~\cite{EM1994,EM1995}. In other words, the condensate $C_{U(1)}$
is chosen to be a {\it genuine} $2L$--fermion condensate, with a zero
``disconnected part'', the latter being the contribution proportional
to $\langle \bar{q} q \rangle^L$,
corresponding to retaining the vacuum intermediate state in all the channels
and neglecting the contributions of all the other states.
As a remark, we observe that the condensate $C_{U(1)}$ so defined
turns out to be of order ${\cal O}(g^{2L - 2} N_c^L) = {\cal O}(N_c)$
in the large--$N_c$ expansion, exactly as the quark condensate
$\langle \bar{q} q \rangle$.

In the case of the $SU(L) \otimes SU(L)$ chiral symmetry, one finds
the following Ward identity (WI):
\be
\int d^4 x \langle T\partial^\mu A^a_\mu (x) 
i\bar{q} \gamma_5 T^b q(0) \rangle
= i \delta _{ab} {1 \over L} \langle \bar{q} q \rangle ,
\label{eqn6}
\ee
where $A^a_\mu = \bar{q}\gamma_\mu \gamma_5 T^a q$ is the $SU(L)$ axial
current. If $\langle \bar{q} q \rangle \ne 0$ (in the chiral limit
$sup(m_i) \to 0$), the anomaly--free WI (\ref{eqn6}) implies the existence
of $L^2-1$ non--singlet NG bosons, interpolated by the hermitean fields
$O_b = i \bar{q} \gamma_5 T^b q$.
Similarly, in the case of the $U(1)$ axial symmetry, one finds that:
\be
\int d^4x \langle T\partial^\mu J_{5, \mu}(x) i\bar{q} \gamma_5 q(0)
\rangle = 2i \langle \bar{q} q \rangle ,
\label{eqn7}
\ee
where $J_{5, \mu} = \bar{q} \gamma_\mu \gamma_5 q$ is the $U(1)$ axial current.
But this is not the whole story! One also derives the following WI:
\be
\int d^4x \langle T\partial^\mu J_{5, \mu}(x) O_P(0) \rangle = 
2Li \langle {\cal L}_{eff} \rangle ,
\label{eqn8}
\ee
where $O_P \sim i[ {\det}(\bar{q}_{sL}q_{tR}) - {\det}(\bar{q}_{sR}q_{tL}) ]$.
If the $U(1)$ chiral symmetry is still broken above $T_{ch}$, i.e.,
$\langle {\cal L}_{eff} \rangle \ne 0$ for $T > T_{ch}$
(while $\langle \bar{q} q \rangle = 0$ for $T > T_{ch}$), then this WI
implies the existence of a (pseudo--)Goldstone boson (in the large--$N_c$
limit!) coming from this breaking and interpolated by the hermitean
field $O_P$. Therefore, the $U(1)_A$ (pseudo--)NG boson (i.e.,
the $\eta'$) is an ``exotic'' $2L$--fermion state for $T > T_{ch}$!

\section{The new chiral effective Lagrangian}

We shall see in this section how the proposed scenario, in which the $U(1)$
axial symmetry is still broken above the chiral transition, can be
consistently reproduced using an effective--Lagrangian model.~\cite{EM1994}

It is well known that the low--energy dynamics of the pseudoscalar mesons, 
including the effects due to the anomaly and the quark condensate, and
expanding to the first order in the light quark masses, can be described,
in the large--$N_c$ limit, by an effective Lagrangian,
\cite{DiVecchia-Veneziano80,Witten80,Rosenzweig-et-al.80,Nath-Arnowitt81}
written in terms of the mesonic field $U_{ij} \sim \bar{q}_{jR} q_{iL}$
(up to a multiplicative constant) and the topological charge density $Q$.
If we make the assumption that the $U(1)$ chiral symmetry is restored at
a temperature $T_{U(1)}$ greater than $T_{ch}$, we need another order parameter
for the $U(1)$ chiral symmetry, the form of which has been 
discussed in the previous section. We must now define a field variable $X$,
associated with this new condensate, to be inserted in the chiral Lagrangian.
The translation from the fundamental quark fields to the
effective--Lagrangian meson fields is done as follows. The operators
$i \bar{q} \gamma_5 q$ and $\bar{q} q$ entering in the WI (\ref{eqn7})
are essentially equal to (up to a multiplicative constant)
$i(\Tr U - \Tr U^\dagger)$ and $\Tr U + \Tr U^\dagger$ respectively.
Similarly, the operators
${\cal L}_{eff} \sim {\det}(\bar{q}_{sL}q_{tR})
+ {\det}(\bar{q}_{sR}q_{tL})$ and
$O_P \sim i[ {\det}(\bar{q}_{sL}q_{tR})
- {\det}(\bar{q}_{sR}q_{tL}) ]$
entering in the WI (\ref{eqn8}) can be put equal to (up to a multiplicative
constant) $X + X^\dagger$ and $i(X - X^\dagger)$ respectively, where
$X \sim {\det} \left( \bar{q}_{sR} q_{tL} \right)$
is the new field variable (up to a multiplicative constant),
related to the new $U(1)$ chiral condensate, which must be inserted
in the chiral effective Lagrangian.
It was shown in Refs.~\cite{EM1994} that the most simple effective 
Lagrangian, constructed with the fields $U$, $X$ and $Q$, is:
\bea
\lefteqn{
{\cal L}(U,U^\dagger ,X,X^\dagger ,Q) =
{1 \over 2}\Tr(\partial_\mu U\partial^\mu U^\dagger )
+ {1 \over 2}\partial_\mu X\partial^\mu X^\dagger + } \nonumber \\
& & -V(U,U^\dagger ,X,X^\dagger ) +{1 \over 2}iQ(x)\omega_1 \Tr[\log(U) - 
\log(U^\dagger )]+ \nonumber \\
& & +{1 \over 2}iQ(x)(1-\omega_1)[\log(X)-\log(X^\dagger )]+{1 \over 2A}Q^2(x),
\label{eqn9}
\eea
where the potential term $V(U,U^{\dagger},X,X^{\dagger})$ has the form:
\bea
\lefteqn{
V(U,U^\dagger ,X,X^\dagger )={1 \over 4}\lambda_{\pi}^2 \Tr[(U^\dagger U
-\rho_\pi 
\cdot {\bf I})^2] +
{1 \over 4}\lambda_X^2 (X^\dagger X-\rho_X )^2 } \nonumber \\
& & -{B_m \over 2\sqrt{2}}\Tr[MU+M^\dagger U^\dagger ]
-{c_1 \over 2\sqrt{2}}[\det(U)X^\dagger + \det(U^\dagger )X].
\label{eqn10}
\eea
$M$ represents the quark mass matrix, $M={\rm diag}(m_1,\ldots,m_L)$,
and $A$ is the topological susceptibility in the pure YM theory.
All the parameters appearing in the Lagrangian are to be considered as 
functions of the physical temperature $T$. In particular, the parameters 
$\rho_{\pi}$ and $\rho_X$ are responsible for the behaviour of the theory 
respectively across the $SU(L) \otimes SU(L)$ 
and the $U(1)$ chiral phase transitions, according to the following table:
$$
\vbox{\tabskip=0pt \offinterlineskip
\halign to320pt{\strut#
& \vrule#\tabskip=1em plus2em& \hfil#&\vrule#
& \hfil#&\vrule#
& \hfil#&\vrule#
& \hfil#&\vrule#
\tabskip=0pt\cr\noalign{\hrule}
&& \omit\hidewidth { } \hidewidth
&& \omit\hidewidth $T<T_{ch}$ \hidewidth
&& \omit\hidewidth $T_{ch}<T<T_{U(1)}$ \hidewidth
&& \omit\hidewidth $T>T_{U(1)}$ \hidewidth& \cr
\noalign{\hrule}
&& { } & & { } & & { } & & { } &\cr
&& $\rho_\pi$ & & ${1\over 2}F_\pi^2>0$ & & $-{1\over 2}B_\pi^2<0$ & 
& $-{1\over 2}B_\pi^2<0$ &\cr 
&& { } & & { } & & { } & & { } &
\cr\noalign{\hrule}
&& { } & & { } & & { } & & { } &\cr
&& $\rho_X$ & & ${1\over 2}F_X^2>0$ & & ${1\over 2}F_X^2>0$ & 
& $-{1\over 2}B_X^2<0$ &\cr 
&& { } & & { } & & { } & & { } &
\cr\noalign{\hrule}\noalign{\smallskip}& \multispan7 [Tab.1]
\hfil\cr}}
$$
[That is: $\rho_{\pi}(T_{ch})=0$ and $\rho_X(T_{U(1)})=0$.]
The $U(1)$ chiral symmetry remains broken also in the region of 
temperatures $T_{ch} < T < T_{U(1)}$, where on the contrary the 
$SU(L) \otimes SU(L)$ chiral symmetry is restored.
The $U(1)$ chiral symmetry is restored above $T_{U(1)}$.
We also assume that the topological susceptibility of the pure YM theory,
$A(T)$, drops to zero at a temperature $T_{\chi}$ greater than $T_{ch}$.
Then, for the consistency of the model, it must be $T_{\chi} \le T_{U(1)}$.

One can study the mass spectrum of the theory for $T < T_{ch}$ and
$T_{ch} < T < T_{U(1)}$. First of all, let us see what happens for
$T<T_{ch}$, where both the $SU(L) \otimes SU(L)$ and the $U(1)$ chiral
symmetry are broken. Integrating out the field variable $Q$ and taking only
the quadratic part of the Lagrangian, one finds that, in the chiral limit
$sup(m_i) \to 0$, there are $L^2-1$ zero--mass states, which represent the
$L^2-1$ Goldstone bosons coming from the breaking of the $SU(L) \otimes SU(L)$
chiral symmetry down to $SU(L)_V$. Then there are two singlet eigenstates:
\bea
\eta' &=& {1 \over \sqrt{F_\pi^2 + LF_X^2}}(\sqrt{L}F_X S_X + F_\pi S_\pi), 
\nonumber \\
\eta_X &=& {1 \over \sqrt{F_\pi^2 + LF_X^2}}(-F_\pi S_X + \sqrt{L}F_X S_\pi), 
\label{eqn11}
\eea
(where $S_\pi$ is the usual $SU(L)$ singlet meson field associated with $U$,
while $S_X$ is the meson field associated with $X$) with non--zero masses.
One of them ($\eta'$) has a ``light'' mass, in the sense of the
$N_c \to \infty$ limit, being
\be
m^2_{\eta'} = {2LA \over F_\pi^2 + LF_X^2} = {\cal O}({1 \over N_c}).
\label{eqn12}
\ee
This mass is intimately related to the anomaly and they both vanish in the 
$N_c \to \infty$ limit. On the contrary the field $\eta_X$ has a sort of
``heavy hadronic'' mass of order ${\cal O}(1)$ in the large--$N_c$ limit.
We immediately see that, if we put $F_X=0$ in the above--written
formulae (i.e., if we neglect the new $U(1)$ chiral condensate), then
$\eta' = S_\pi$ and $m^2_{\eta'}$ reduces to ${2LA \over F^2_\pi}$, which is
the ``usual'' $\eta'$ mass in the chiral limit.~\cite{Witten79,Veneziano79}
Yet, in the general case $F_X \ne 0$, the two states which diagonalize the 
squared mass matrix are linear combinations of the ``quark--anti-quark'' 
singlet field $S_\pi$ and of the ``exotic'' field $S_X$.
Both the $\eta'$ and the $\eta_X$ have the same quantum numbers (spin, 
parity and so on), but they have a different quark content: one is mostly
$\sim i(\bar{q}_{L}q_{R}-\bar{q}_{R}q_{L})$, while the other is mostly
$\sim i[ {\det}(\bar{q}_{sL}q_{tR}) - {\det}(\bar{q}_{sR}q_{tL}) ]$.
What happens when approaching the chiral transition temperature $T_{ch}$?
We know that $F_\pi(T) \to 0$ when $T \to T_{ch}$. From Eq. (\ref{eqn12})
we see that $m^2_{\eta'}(T_{ch}) = {2A \over F_X^2}$
and, from the first Eq. (\ref{eqn11}), $\eta'(T_{ch}) = S_X$.
We have continuity in the mass spectrum of the theory through the chiral 
phase transition at $T=T_{ch}$.
In fact, if we study the mass spectrum of the theory in the region of
temperatures $T_{ch} < T < T_{U(1)}$ (where the $SU(L) \otimes SU(L)$ chiral
symmetry is restored, while the $U(1)$ chiral symmetry is still broken),
one finds that there is a singlet meson field $S_X$ (associated with the 
field $X$ in the chiral Lagrangian) with a squared mass given by (in the
chiral limit): $m^2_{S_X} = {2A \over F_X^2}$.
This is nothing but the {\it would--be} Goldstone particle 
coming from the breaking of the $U(1)$ chiral symmetry, i.e., the $\eta'$,
which, for $T>T_{ch}$, is a sort of ``exotic'' matter field of the form
$\sim i[ {\det}(\bar{q}_{sL}q_{tR})
- {\det}(\bar{q}_{sR}q_{tL}) ]$.
Its existence could be proved perhaps in the near
future by heavy--ions experiments. 

\section{Conclusions}

We have tried to gain a physical insight into the breaking mechanism of the
$U(1)$ axial symmetry, through a study of the behaviour of the theory at
finite temperature. In the following, we briefly summarize the main points
that we have discussed and the results that we have obtained.

\begin{itemize}
\item{
One expects that, above a certain critical temperature, also the $U(1)$ 
axial symmetry will be (approximately) restored. We have tried to see if
this transition has (or has not) anything to do with the usual
$SU(L) \otimes SU(L)$ chiral transition: various possible scenarios
have been discussed.
}
\item{
We have proposed a scenario (supported by lattice results) in which the
$U(1)$ axial symmetry is still broken above the chiral transition.
A new order parameter is introduced for the $U(1)$ axial symmetry.
}
\item{
We have shown that this scenario can be 
consistently reproduced using an effective Lagrangian model.
We have analysed the effects that one should observe on the mass 
spectrum of the theory, both below and above $T_{ch}$. In this scenario,
the $\eta'$ survives across the chiral transition at $T_{ch}$ in the form of
an ``exotic'' $2L$--fermion state.
}
\end{itemize}
This scenario could perhaps be verified in the near future by heavy--ions 
experiments, by analysing the spectrum in the singlet sector.
Some tests and verifications of this picture could also be provided by Monte
Carlo simulations on the lattice.

\end{document}